\documentclass[prl,twocolumn,aps,showpacs,10pt,floatfix,longbibliography]{revtex4-1}

\usepackage[pdftex]{graphicx} 
\usepackage{epstopdf}
\usepackage{verbatim}
\usepackage{color}
\usepackage{subfigure}
\usepackage{tabularx}
\usepackage{amsfonts}
\usepackage{wasysym}
\usepackage{bm}

\newcommand{\avg}[1]{\langle\,#1\,\rangle}

\newcolumntype{C}[1]{>{\centering\let\newline\\\arraybackslash\hspace{0pt}}m{#1}}

\begin{document}
\title{Hidden multipolar orders of dipole-octupole doublets on a triangular lattice}
\author{Yao-Dong Li$^{1}$}
\author{Xiaoqun Wang$^{2,4}$}
\author{Gang Chen$^{1,3,4}$}
\email{gangchen.physics@gmail.com}
\affiliation{$^{1}$State Key Laboratory of Surface Physics,
Center for Field Theory and Particle Physics, Department of Physics,
Fudan University, Shanghai 200433, People's Republic of China}
\affiliation{$^{2}$Department of Physics and Astronomy, 
Shanghai Jiao Tong University, Shanghai 200240, People’s Republic of China}
\affiliation{$^{3}$Perimeter Institute for Theoretical Physics, Waterloo, Ontario N2L 2Y5, Canada}
\affiliation{$^{4}$Collaborative Innovation Center of Advanced Microstructures,
Nanjing, 210093, People's Republic of China}

\date{\today}

\begin{abstract}
Motivated by the recent development in strong spin-orbit-coupled materials,
we consider the dipole-octupole doublets on the triangular lattice. We
propose the most general interaction between these unusual
local moments. Due to the spin-orbit entanglement and the special form of its
wavefunction, the dipole-octupole doublet has a rather peculiar property under
the lattice symmetry operation. As a result, the interaction
is highly anisotropic in the pseudospin space, but remarkably, is uniform 
spatially. We analyze the ground state properties of this generic model and
emphasize the hidden multipolar orders that emerge from the dipolar
and octupolar interactions. We clarify the quantum mutual
modulations between the dipolar and octupolar orders.
We predict the experimental consequences of the multipolar orders 
and propose the rare-earth triangular materials as candidate systems 
for these unusual properties.
\end{abstract}

\maketitle

\emph{Introduction.}---In recent years, there has been an
intensive interest in exploring electron systems
that involve both strong spin-orbit coupling (SOC) and
substantial electron correlations, especially in materials
with heavy elements such as $5d$ transition metal elements
and $4f$ rare-earth elements~\cite{WCKB,rau2016spin,pesin2010mott,savary2016quantum}. 
Because of the spatial orientation of the orbitals, 
the spin-orbit entanglement in strongly correlated Mott insulators 
often gives rise to rather complicated models that involve both spatial 
and spin anisotropies~\cite{Jackeli2009,rau2014generic,rau2016spin,Chen2008,PhysRevB.82.174440,Chen2011}. While this is true for most spin-orbit-entangled moment, 
in this Letter, we propose a remarkably simple model
for a peculiar spin-orbit-entangled doublet, namely
``dipole-octupole doublet'' (DO doublet)~\cite{Huang2014,li2016octupolar}
on a triangular lattice, and connect this model 
with the rare-earth based triangular lattice materials.
Due to the multipolar nature of the interaction,
this simple but realistic model, in a large parameter regime,
realizes {\sl hidden} magnetic multipolar orders and leads to
unexpected experimental consequences.

The search for hidden order is an active field in the $f$ 
electron systems~\cite{RevModPhys.81.807}. 
The magnetic multipolar order has been proposed for 
URu$_2$Si$_2$ and NpO$_2$, and various experimental 
evidence has been found~\cite{RevModPhys.81.807,chandra2013hastatic,chandra2015hastatic}. 
Nevertheless, the precise nature of the multipolar orders 
in URu$_2$Si$_2$ and NpO$_2$ has not come to a consensus. 
It is partly because the complication and multitudes of the degree of freedom
often prohibit the precise modeling of the multipolar interactions in these systems.  
In contrast, our model is a precise modeling of the multipolar interactions 
for the DO doublet systems and might be the simplest such model in 
the strong spin-orbit-coupled materials that realizes hidden multipolar
orders~\cite{WCKB,RevModPhys.81.807}.

\begin{figure}[tp]
\centering
\includegraphics[width=.23\textwidth]{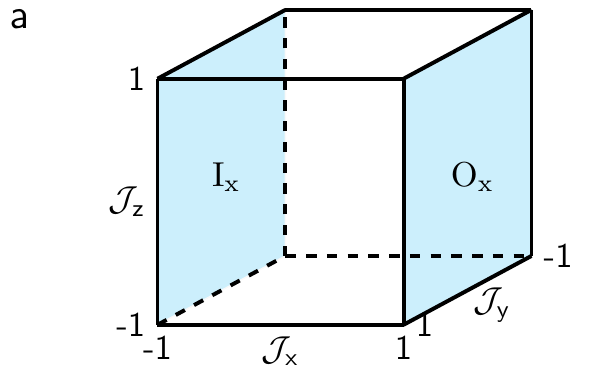}
\includegraphics[width=.23\textwidth]{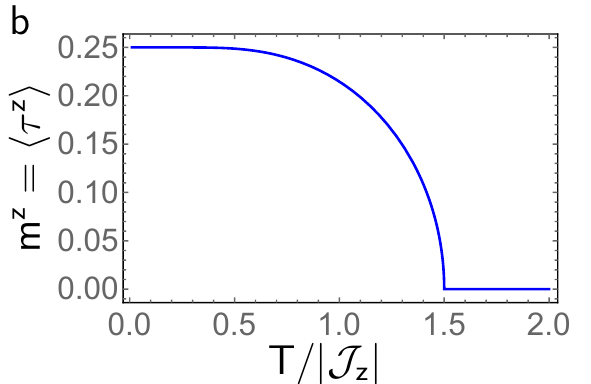}
\includegraphics[width=.23\textwidth]{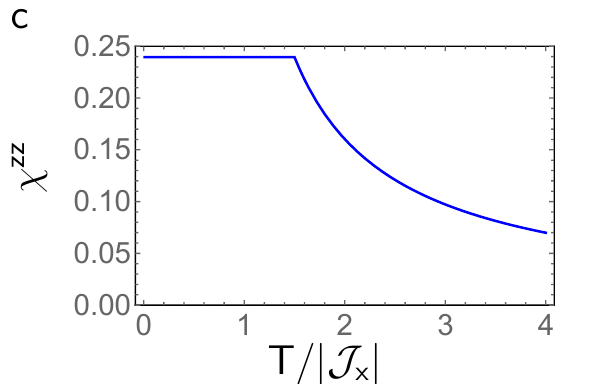}
\includegraphics[width=.23\textwidth]{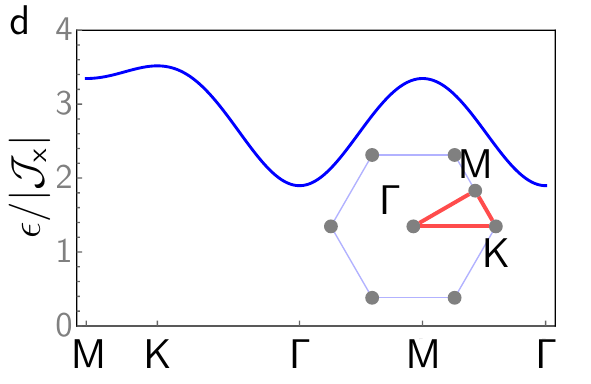}
\caption{
(a) The six surfaces of the cuboid. I$_{\mu}$ (O$_{\mu}$)
refers to the inner (outer) surface with ${\mathcal J}_{\mu} = -1$
(${\mathcal J}_{\mu} = 1$). We have marked the
I$_{\text x}$ and O$_{\text x}$ surfaces.
(b) The magnetization of the FD$_{\text z}$ state
on the I$_{\text z}$ surface with $(\mathcal{J}_{x},
\mathcal{J}_y, \mathcal{J}_z) = (-0.5,-0.2,-1)$
and $\theta = {\pi}/{3}$. The FD$_{\text z}$ transition is
at $T_{\text{d}} = {1.5|\mathcal{J}_{z}|}$.
(c) Magnetic susceptibility $\chi^{zz}$ of the FO state 
on the I$_{\text x}$ surface with $(\mathcal{J}_{x}, 
\mathcal{J}_y, \mathcal{J}_z) = (-1,-0.2,-0.5)$
and $\theta = \pi/3$. The FO transition is at 
$T_{\text{o}} = {1.5|\mathcal{J}_{x}|}$.
(d) Octupolar-wave excitation with the same parameters as in (c).}
\label{fig1}
\end{figure}

Three families of triangular lattice materials, MgYbGaO$_4$~\cite{Yuesheng2015,YaodongPRB,ShenYao201607,Martin201607,YueshengmuSR,Yaodong201608},
the isostructural ternary family RCd$_3$P$_3$, RZn$_3$P$_3$, 
RCd$_3$As$_3$, RZn$_3$As$_3$ (R = Ce, Pr, Nd, Sm)~\cite{Stoyko2011,Nientiedt1999,Yamada2010},
and R$_2$O$_2$CO$_3$ (R = Nd, Sm, Dy)~\cite{Arjun2016}, 
have recently been discovered. These materials
contain the rare-earth elements, whose $4f$ electrons 
involve strong SOC and strong correlations.
The strong SOC entangles the total electron spin ${\bf S}$
with the orbital angular momentum ${\bf L}$ and leads to a
total moment ${\bf J}$. Like the case in the rare-earth 
pyrochlores~\cite{Ross2011},
the local D$_{\text{3d}}$ crystal electric field (CEF) splits the $(2J+1)$ states into the
crystal field states~\cite{YaodongPRB}. For a half-integer 
(integer) $J$, the CEF ground state is a Kramers' doublet 
(either a singlet or a non-Kramers' doublet).
The ground state doublets define the low-temperature 
magnetic properties of the system. In the previous work, 
we proposed a generic anisotropic spin model for 
non-Kramers' doublets and the usual Kramers' doublets on 
a triangular lattice~\cite{YaodongPRB}.
Here we introduce a generic model for the DO Kramers'
doublet on the triangular lattice~\cite{Yuesheng2015,YaodongPRB}
and predict the experimental consequences 
of the hidden multipolar orders.

\emph{Dipole-octupole doublet.}---The DO doublet is a special type of
Kramers' doublet. It occurs when the crystal field ground state wavefunctions
$|\Psi_{\pm}\rangle$ are linear superpositions of the states with
$J^z = 3n/2$ where $n$ is an odd integer. Unlike the usual Kramers doublets
that transform as a two-dimensional irreducible representation of the
D$_{\text{3d}}$ point group~\cite{YaodongPRB}, each state of the DO doublet transforms as 
a one-dimensional irreducible representation ($\Gamma_5^+$ or $\Gamma_6^+$)
of the D$_{\text{3d}}$ point group~\cite{Huang2014}. 
This crucial difference is most easy to be understood if one applies 
the 3-fold rotation along the $z$ axis to these states.
Under the 3-fold rotation, we have $\exp (-i \frac{2\pi}{3} J^z ) \,
|{J^z = {3n}/{2}} \rangle = - |{J^z = {3n}/{2}} \rangle$.
Therefore, the wavefunctions of the DO doublet, $| \Psi_{\pm} \rangle$,
stay invariant under this rotation except getting an overall minus sign,
{\sl i.e.},
\begin{equation}
\exp (-i \frac{2\pi}{3} J^z ) \,
| {\Psi_{\pm}} \rangle = - |{ \Psi_{\pm} } \rangle .
\end{equation}
In contrast, for the usual Kramers' doublet, the two states
would mix with each other under this rotation. 
The degeneracy of the DO doublet is protected by 
time reversal symmetry that switches the two states. 
This special doublet has been found in various neodymium (Nd)
pyrochlores~\cite{NdHfO,NdZrO1,NdSnO,NdZrO2,NdZrO3,NdZrO4,CdYbS}, 
dysprosium (Dy) pyrochlore~\cite{0953-8984-24-25-256003}, 
osmium (Os) pyrochlore~\cite{CdOsO,PhysRevLett.108.247205},
erbium (Er) and ytterbium (Yb) spinels~\cite{CdErO,CdYbS}, 
and Ce$_2$Sn$_2$O$_7$~\cite{CeSnO2015}.
We expect the DO doublet should occur in some of the rare-earth
triangular materials, especially since these rare-earth ions 
experience the same D$_{\text{3d}}$ crystal field environment.

\emph{Generic pseudospin model on a triangular lattice.}---Here we
explain the interaction between the DO doublets on a
triangular lattice. Due to the two-fold degeneracy of the
DO doublet, we introduce the pseudospin operators that 
act on this DO doublet,
$\tau^+ =
{| { \Psi_{+}^{\phantom\dagger} } \rangle} \langle{  \Psi_{-}^{\phantom\dagger} } |$,
$\tau^- =
{| { \Psi_{-}^{\phantom\dagger} } \rangle} \langle{ { \Psi_{+}^{\phantom\dagger} } } |$,
$\tau^z =
\frac{1}{2}{| { \Psi_{+}^{\phantom\dagger} } \rangle} \langle{ { \Psi_{+}^{\phantom\dagger} } } |
-
\frac{1}{2}{| { \Psi_{-}^{\phantom\dagger} } \rangle} \langle{ { \Psi_{-}^{\phantom\dagger} } } |
$, where $\tau^{\pm} \equiv \tau^x \pm i \tau^y$. To obtain the
exchange interaction, we start with the symmetry properties of 
the pseudospins under the space group symmetry.

For all the three families of rare-earth triangular 
lattice materials~\cite{Yuesheng2015,YaodongPRB,ShenYao201607,Martin201607,
YueshengmuSR,Yaodong201608,Stoyko2011,Nientiedt1999,Yamada2010,Arjun2016}, 
the space group is either R$\bar{3}$m or P6$_3$mmc. 
As all rare-earth ions in these materials have a layered 
triangular structure and the interlayer separation is much
larger than the intralayer lattice constant, it is 
sufficient to just keep the interaction within the triangular layer
and ignore the interlayer couplings. As far as the space group
symmetry is concerned, we only need to retain the symmetry generators
that operate within each triangular layer.
It turns out that, for a single triangular layer,
both R$\bar{3}$m and P6$_3$mmc space groups give
a three-fold rotation around the $z$ axis, $C_3$,
a two-fold rotation about the diagonal direction, $C_2$,
a site inversion symmetry $I$, and two lattice translations,
$T_x$ and $T_y$. The symmetry operation on
$\tau^{\mu}_{\bf r}$ is given as~\cite{Supple}
\begin{eqnarray}
\left\{
\begin{array}{llll}
C_3: & \tau^{x}_{\bf r} \rightarrow \tau^{x}_{ C_3 ({\bf r}) }, &
\tau^{y}_{\bf r} \rightarrow \tau^{y}_{ C_3 ({\bf r}) }, &
\tau^{z}_{\bf r} \rightarrow \tau^{z}_{ C_3 ({\bf r}) },
\\
C_2: &\tau^{x}_{\bf r} \rightarrow  \tau^{x}_{ C_2 ({\bf r}) }, &
\tau^{y}_{\bf r} \rightarrow  -\tau^{y}_{ C_2 ({\bf r}) }, &
\tau^{z}_{\bf r} \rightarrow  -\tau^{z}_{ C_2 ({\bf r}) } ,
\\
I: & \tau^{x}_{\bf r} \rightarrow \tau^{x}_{ I ({\bf r}) }, &
\tau^{y}_{\bf r} \rightarrow \tau^{y}_{ I ({\bf r}) }, &
\tau^{z}_{\bf r} \rightarrow \tau^{z}_{ I ({\bf r}) },
\\
T_x: & \tau^{x}_{\bf r} \rightarrow \tau^{x}_{ T_{x} ({\bf r}) }, &
\tau^{y}_{\bf r} \rightarrow \tau^{y}_{ T_{x} ({\bf r}) }, &
\tau^{z}_{\bf r} \rightarrow \tau^{z}_{ T_{x} ({\bf r}) },
\\
T_y: & \tau^{x}_{\bf r} \rightarrow \tau^{x}_{ T_{y} ({\bf r}) }, &
\tau^{y}_{\bf r} \rightarrow \tau^{y}_{ T_{y} ({\bf r}) }, &
\tau^{z}_{\bf r} \rightarrow \tau^{z}_{ T_{y} ({\bf r}) }.
\end{array}
\right.
\label{symmetry}
\end{eqnarray}
Since the $4f$ electron wavefunction is very localized,
we only need to keep the nearest-neighbor interactions.
The most general nearest-neighbor model, 
allowed by the above symmetries,
is given as
\begin{eqnarray}
H_0 &=& \sum_{\langle {\bf r}{\bf r}' \rangle}
\big[J_x^{\phantom\dagger} \tau^x_{\bf r} \tau^x_{{\bf r}'}
+ J_y^{\phantom\dagger} \tau^y_{\bf r} \tau^y_{{\bf r}'}
+ J_z^{\phantom\dagger} \tau^z_{\bf r} \tau^z_{{\bf r}'}
\nonumber \\
&&
+ J_{yz}^{\phantom\dagger}
(\tau^y_{\bf r} \tau^z_{{\bf r}'} + \tau^z_{\bf r} \tau^y_{{\bf r}'})
\big].
\label{model}
\end{eqnarray}
Here we give a few comments on this model. First of all,
the pseudospin interaction is anisotropic in the pseudospin space
because of the spin-orbit entanglement in the DO doublet. 
What is surprising is that the interaction is
spatially uniform and is identical for every bond orientation.
This is unusual since the orbitals have orientations.
This remarkable spatial property comes from the peculiar
symmetry property of the DO doublet in Eq.~(\ref{symmetry}).
Secondly, there exists a crossing coupling between $\tau^y$
and $\tau^z$ because $\tau^y$ and $\tau^z$ transform identically and
behave like the magnetic dipole moments under the space group.
Thirdly, there is no crossing coupling between $\tau^x$
and $\tau^y$ or $\tau^z$ because $\tau^x$ transforms
as an octupole moment under the space group.
This holds even for further neighbor interactions~\cite{GangChenunpublished}.
The $J_x$ interaction is the interaction between the octupole moments.

Another remarkable property of the DO doublet
is the infinite anisotropy in the Land\'{e} $g$-factor
when it couples to an external magnetic field.
After including the Zeeman term, we have the full Hamiltonian
$H = H_0 - h  \sum_{\bf r} \tau^z_{\bf r}$.
Due to the spatial uniformity of the interaction,
we are able to implement a rotation by an angle $\theta$
around the $x$ direction in the pseudospin space
and eliminate the crossing coupling between 
$\tau^y$ and $\tau^z$. The reduced model is given as
\begin{eqnarray}
H &=& \sum_{\langle {\bf r}{\bf r}' \rangle}
\big[{\mathcal J}_x^{\phantom\dagger} T^x_{\bf r} T^x_{{\bf r}'}
+ {\mathcal J}_y^{\phantom\dagger} T^y_{\bf r} T^y_{{\bf r}'}
+ {\mathcal J}_z^{\phantom\dagger} T^z_{\bf r} T^z_{{\bf r}'}  ]
\nonumber \\
&& - h \sum_{\bf r} [\cos \theta \, T^z_{\bf r} + \sin \theta \, T^y_{\bf r}],
\label{newmodel}
\end{eqnarray}
where $T^x = \tau^x, T^y = \tau^z \sin \theta + \tau^y \cos \theta,
T^z= \tau^z \cos \theta  - \tau^y \sin \theta$,
and ${\mathcal J}_x$, ${\mathcal J}_y$, ${\mathcal J}_z$ are defined in
the Supplementary information. Note both $T^y$ and $T^z$ behave 
like dipole moments. Like the XYZ model on the pyrochlore lattice~\cite{Huang2014,li2016octupolar},
this model does not have a sign problem for
quantum Monte Carlo simulation in a large parameter regime,
and this is valid on any other lattices such as the 3D
FCC lattice where DO doublets could exist~\cite{PhysRevB.82.174440}.

\emph{Hidden ferro-octupolar orders.}---We now explain the
hidden multipolar orders of the model in Eq.~(\ref{newmodel}).
We start with the parameter regime on the I$_{\text z}$ surface with
$\mathcal{J}_z = -1$ (see Fig.~\ref{fig1}a).
This regime simply gives a conventional ferromagnetic 
 ground state with a uniform $\langle T^z \rangle$.  
Since $T^z$ is a dipole moment, this state is dubbed ferro-dipolar 
(FD$_{\text z}$) state, where the subindex $z$ refers to the 
direction of the dipole moment. With a ferromagnetic dipole 
moment, this state can be readily confirmed in a magnetization 
measurement.

The reduced model in Eq.~(\ref{newmodel}) has an interesting
permutation structure. Using the result on the I$_{\text z}$ surface,
we can generate the ground states on the I$_{\text y}$ surface
with $\mathcal{J}_y = -1$ and the I$_{\text x}$ surface with
$\mathcal{J}_x = -1$. As the FD$_{\text y}$ order of the I$_{\text y}$ surface
shares the same symmetry as the FD$_{\text z}$ order of the I$_{\text z}$ surface,
we do not give a repeated discussion here.
Although the permutation trick to relate different regimes seems simple,
the physics on the I$_{\text x}$ surface is rather special and unconventional,
and it is this distinction that we clarify below.
Clearly, as $\langle T^x \rangle$ is uniform and non-zero
on the I$_{\text x}$ surface, time reversal symmetry is explicitly
broken and the ground state is a ferromagnetic state.
As we compute within the mean-field theory in 
the Supplementary Information and show in Fig.~\ref{fig1}, 
however, the magnetic susceptibility does not show any divergent 
behavior. This is very different
from what we would naively expect for an usual ferromagnetic state.
The order parameter $\langle T^x \rangle$ is an octupole moment
and does not couple linearly to the external magnetic field.
Therefore, it is hidden in the usual magnetization measurement.

\begin{figure}[t]
\centering
\includegraphics[width=6.5cm]{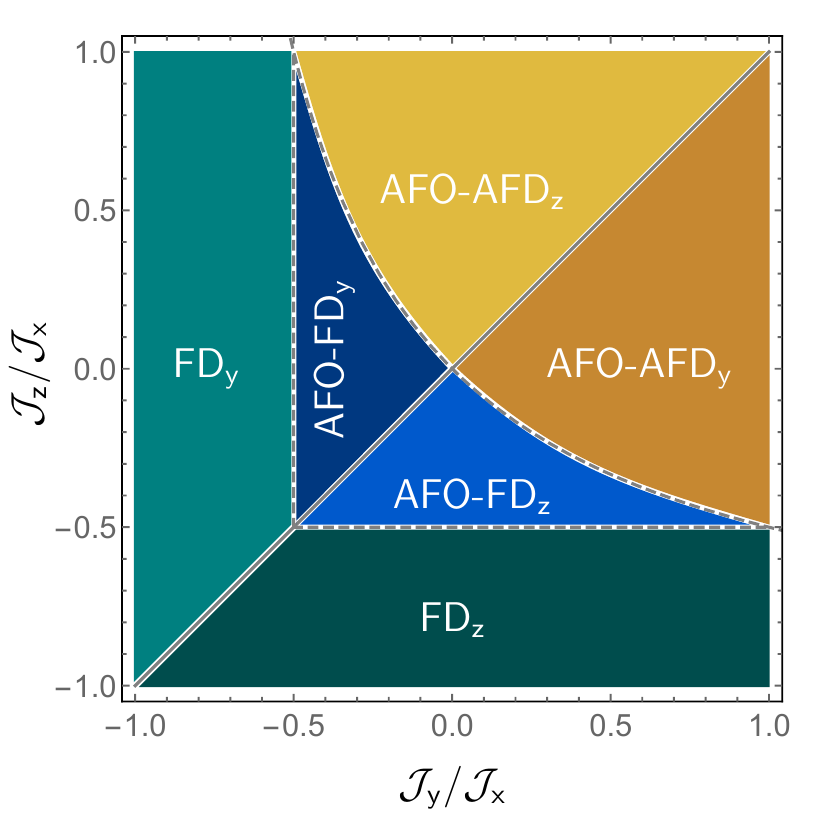}
\caption{(Color online.) The phase diagram on the O$_{\text x}$
surface ($\mathcal{J}_x = 1$). Solid (dashed) lines
indicate first (continuous) order phase transitions.
}
\label{fig2}
\end{figure}

Despite its invisibility in the usual thermodynamic measurements,
one could instead search for the evidence of the octupolar order
by other experimental probes. Since the octupolar order
explicitly breaks time reversal symmetry, polar Kerr effect
could be used to detect the time reversal symmetry breaking~\cite{Cho2016}.
Moreover, inside the FO phase, the dipole moment $\tau^z$ 
flips the octupole moment and creates octupolar-wave
excitations. As $\tau^z$ directly couples to the neutron spin,
the octupolar-wave excitation can be directly detected by
an inelastic neutron scattering experiment. Using the Holstein-Primakoff
boson transformation~\cite{Supple}, we obtain the octupolar-wave dispersion,
\begin{eqnarray}
\omega_{\bf k} &=&  \big[\mathcal{J}_y \sum_{i} \cos{[{\bf k}\cdot {\bf a}_i]}
-3 \mathcal{J}_x \big]^{\frac{1}{2}}
\nonumber \\
& \times & \big[ \mathcal{J}_z \sum_{i} \cos{[{\bf k}\cdot {\bf a}_i]}
-3 \mathcal{J}_x
\big]^{\frac{1}{2}},
\label{ferro-magnon}
\end{eqnarray}
where the summation is over the three nearest neighboring vectors
${\bf a}_1 = (1,0)$, ${\bf a}_2 = (-{1}/{2},{\sqrt{3}}/{2})$,
and ${\bf a}_3 = (-{1}/{2},-{\sqrt{3}}/{2})$.
One should observe a well-defined octupolar wave excitation
below the FO transition despite the absence of
ordering in the magnetization measurement.
This mode is generically gapped because of the low
symmetry of the model. We depict the octupolar wave excitation
in Fig.~\ref{fig1}d.

\emph{Hidden antiferro-octupolar orders.}---Here we consider the
parameter regimes where the dominant interaction is antiferromagnetic.
We focus on the O$_{\text x}$ surface where the
octupolar exchange coupling ${\mathcal J}_x$
is antiferromagnetic and dominant. 
For the O$_{\text y}$ and the O$_{\text z}$ surfaces,
one can apply the permutation on the O$_{\text x}$ surface
and generate the phase diagrams and the relevant phases .
In the absence of the exchange couplings ${\mathcal J}_y$ and ${\mathcal J}_z$,
the Ising exchange interaction ${\mathcal J}_x$ 
is highly frustrated on the triangular lattice.
Any state that satisfies the ``2-plus 1-minus'' or ``2-minus 1-plus''
condition for the $T^x$ configuration on every triangle
is the ground state. Therefore, the ground state is extensively degenerate.

In the XXZ limit of the model with ${\mathcal J}_y = {\mathcal J}_z$, the
weak $\mathcal{J}_y$ and $\mathcal{J}_z$ exchanges allows the system
to tunnel quantum mechanically within the degenerate ground state
manifold and lifts the degeneracy via an order by quantum disorder 
effect~\cite{PhysRevLett.95.127207,PhysRevLett.95.127205,PhysRevLett.112.127203,PhysRevB.91.081104}.
It is well established that the system develops a supersolid order
in a large parameter regime of the XXZ limit. 
With a supersolid order, the system spontaneously breaks the 
U(1) symmetry with $\langle T^{y,z} \rangle \neq 0$
and the translation symmetry with $\langle T^x\rangle \neq 0$.
Moreover, the system has a 3-sublattice magnetic structure
in the supersolid phase. 

To obtain the phase diagram away
from the XXZ limit, we implement a self-consistent
mean-field theory by assuming a 3-sublattice structure for the
mean-field ansatz~\cite{Supple}. Via the mean-field decoupling, 
we have
\begin{eqnarray}
H_{\rm MF} & = &
3 \sum_{{\bf r} \in {\text A} }
\sum_\mu \big[ \mathcal{J}_\mu (m_{\text B}^\mu + m_{\text C}^\mu)
\, T_{\bf r}^\mu \big]
\nonumber \\
& + & 3 \sum_{{\bf r} \in {\text B}}
\sum_\mu \big[ \mathcal{J}_\mu (m_{\text C}^\mu + m_{\text A} ^\mu)
\, T_{\bf r}^\mu \big]
\nonumber \\
& + & 3 \sum_{{\bf r} \in {\text C} }
\sum_\mu \big[ \mathcal{J}_\mu (m_{\text A}^\mu + m_{\text B}^\mu)
\, T_{\bf r}^\mu \big]
\nonumber \\
& - &
h \sum_{\bf r} \big[ {\cos \theta \, T^z_{\bf r} + \sin \theta \, T^y_{\bf r} } \big],
\end{eqnarray}
where $m^\mu_{\Lambda} = \langle T^\mu_{\bf r} \rangle $ is determined
self-consistently for ${\bf r} \in \Lambda$-th sublattice with $\Lambda
= \text{A, B, C}$. Such a mean-field theory captures both the uniform state
and the 3-sublattice state.
The mean-field phase diagram is depicted in Fig.~\ref{fig2}.
The FD$\rm{}_y$ and the FD$\rm{}_z$ phases are the previously mentioned
ferro-dipolar orders with an uniform $\langle T^y \rangle \neq 0$
and $\langle T^z \rangle \neq 0$, respectively.
There is no octupolar order here. It is the considerable ferro-dipolar
interaction in these regions that competes with the antiferro-octupolar
interaction and competely suppresses any octupolar order.

\begin{figure}[tp]
\centering
\includegraphics[height=.167\textwidth]{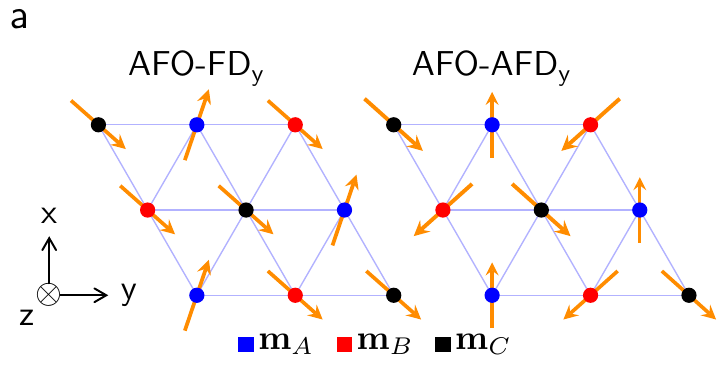}
\includegraphics[height=.167\textwidth]{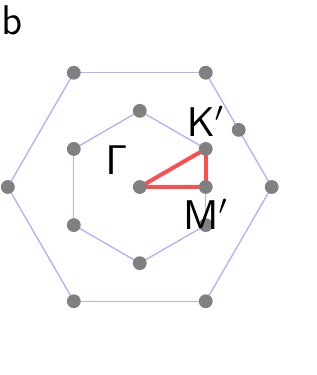}
\includegraphics[width=.23\textwidth]{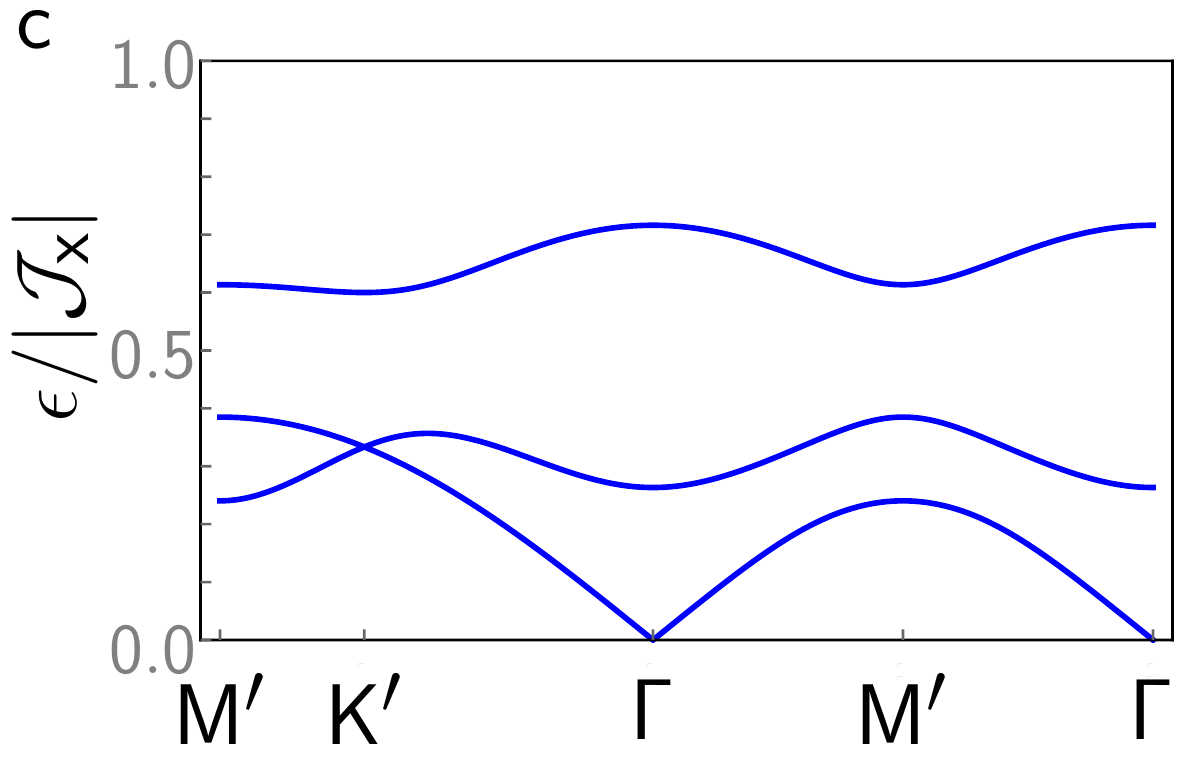}
\includegraphics[width=.23\textwidth]{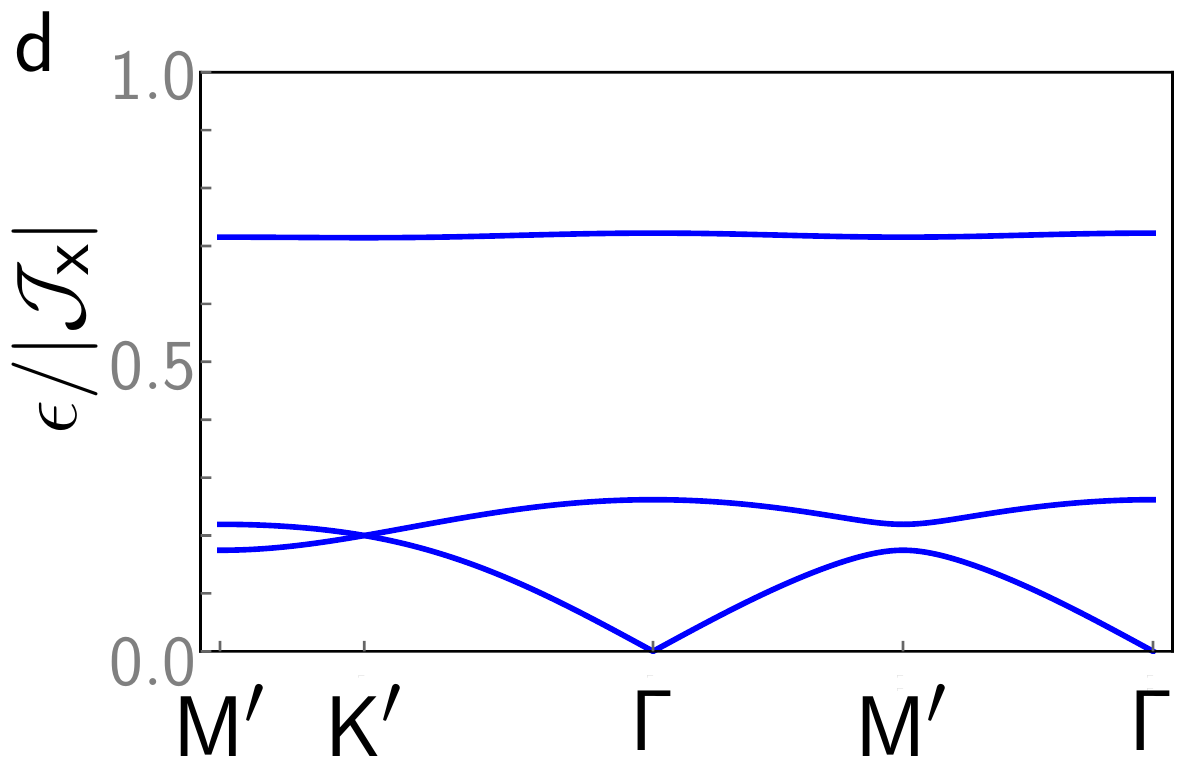}
\caption{
(a) The ordering pattern in the AFO-FD${}_{\rm y}$ and AFO-AFD${}_{\rm y}$ 
phases. The local moments are in the $xy$-plane of the pseudospin space.
The pseudospin configurations of phases across the diagonal line of the 
phase diagram in Fig.~\ref{fig2} are related by interchanging $y$ and $z$
components of ${\bf m}$. Inset is the coordinate system for the pseudospins. 
(b) The original Brillouin zone and the folded Brillouin 
zone due to the 3-sublattice ordering. 
(c, d) Excitation spectrum by linear spin-wave theory with
dominant antiferromagnetic $\mathcal{J}_x$,
for (c) $(\mathcal{J}_x, \mathcal{J}_y, \mathcal{J}_z) = (1,0.4,0)$
in AFO-AFD${}_{\rm y}$, and
(d) $(\mathcal{J}_x, \mathcal{J}_y, \mathcal{J}_z) = (1,-0.4,0)$
in AFO-FD${}_{\rm y}$.}
\label{fig3}
\end{figure}

In region AFO-FD$\rm{}_y$ (AFO-FD$\rm{}_z$) where the
transverse exchange $\mathcal{J}_{ y}$ ($\mathcal{J}_{ z}$) is reduced,
the octupole moment $T^x$ orders antiferromagnetically and
develops a 3-sublattice structure while the dipole moment
$T^y$ ($T^z$) remains ferromagnetically ordered (see Fig.~\ref{fig3}a). Therefore,
the phase is listed as AFO-FD$\rm{}_y$ (AFO-FD$\rm{}_z$).
In these regions, the weak ferro-dipolar interaction allows
the system to fluctuate within the extensively degenerate
ground state manifold of the predominant antiferro-octupolar
interaction and breaks the degeneracy, leading
to the 3-sublattice octupolar order.
The background 3-sublattice octupolar order further modulates
the ferro-dipolar order and renders the 3-sublattice structure
to the ferro-dipolar order. Such a mutual modulation between
unfrustrated ferro-dipolar and the frustrated antiferro-octupolar
interactions is in fact a {\sl quantum effect},
and cannot occur in a classical spin system with the same model.

The 3-sublattice structure of the ferro-dipolar order is a
direct consequence of the underlying antiferromagnetic
octupolar order. This 3-sublattice structure,
however, is completely hidden in the magnetization measurement
that merely gives a finite net magnetization.
To reveal the underlying 3-sublattice structure, one would need
local probes such as NMR and $\mu$SR. The nuclear spin and muon
spin only couple to the dipolar moment, and probe the
local dipolar orders of different sublattices. Alternatively,
the elastic neutron scattering directly probes the
structure of the dipolar orders, and would
observe the magnetic Bragg peaks at
the $\Gamma$ point that corresponds to the
uniform part of the dipolar order
as well as the K points that correspond to the
3-sublattice modulation of the dipolar order.
Besides the static properties, the system supports
three bands of excitations because of the 3-sublattice
structure of the octupolar order. This can be well-observed
in an inelastic neutron scattering measurement.
We plot the the magnetic excitations in Fig.~\ref{fig3}c.

In region AFO-AFD$\rm{}_y$ (AFO-AFD$\rm{}_z$), 
the transverse coupling $\mathcal{J}_{y}$ ($\mathcal{J}_{z}$)
is antiferromagnetic. The system is therefore frustrated, 
and due to frustration the 3-sublattice structure persists for rather
large $\mathcal{J}_{y}$ and $\mathcal{J}_{z}$. Besides the 
antiferromagnetic order of the octupolar moment
$T^x$, the dipolar moments are also antiferromagnetically ordered 
(Fig.~\ref{fig3}a). The ordering of the local moments is constrained 
to either $xy$- or $xz$-plane depending on the magnitude of 
$\mathcal{J}_{y}$ and $\mathcal{J}_{z}$, as in AFO-FD$\rm{}_y$ 
and AFO-AFD$\rm{}_z$ phases. The net magnetization of the dipolar 
moments in AFO-AFD phases is always zero, hence hidden to the
thermodynamic measurements;  but the 3-sublattice structure can manifest itself 
in the spin-wave excitations with 3 bands (see Fig.~\ref{fig3}d). 
The gapless modes at $\Gamma$ in Fig.~\ref{fig3}c and d 
are accidentical due to the extended degeneracy in the Ising limit
and should be gapped when the magnon interactions 
are included.

\emph{Discussion.}---It has been realized that a strong SOC could create
a significant interaction between the magnetic multipole moments.
The magnetic multipolar orders have been proposed in several
strong spin-orbit-coupled systems, e.g. the quadrupolar orders
and the octupolar orderes in ordered double perovskites~\cite{PhysRevB.82.174440}.
The magnetic dipolar orders, being time reversally odd,
are often concomitant with the magnetic octupolar orders.
Since the former plays a dominant role in many magnetic measurements, 
it could complicate the interpretation of many experiments and
the identification of the underlying octupolar orders.
For the DO doublet on the triangular lattice, the lattice symmetry
naturally distinguishes the octupole moments from the dipole ones
and allows them to have independent structures.


The peculiar property of the DO doublets arises from
the wavefunction, and has little to do with the value
of the total moment $J$. Any moment with $J>1/2$ can potentially
support a DO doublet as the CEF ground state doublet.
There is no need to restrict $J$ to be odd integer multiples of $3/2$.
It gives a lot more room for the experimental discovery
of DO doublets in the rare-earth triangular lattice materials.
The experimental studies of the rare-earth
triangular lattice materials have just started.
The CEF ground states of most magnetic ions have not
been understood. A systematic study of the CEFs will be of great interest. 
The magnetic properties of many materials in these families 
are not yet known, and a careful experimental investigation is highly needed.

To summarize, we propose a peculiar Kramers' doublet, namely,
the dipole-octupole doublet, on a triangular lattice. We
propose a rather simple model to describe the 
interaction between the dipole-octupole doublets
and predict the hidden magnetic multipolar order and 
various unexpected properties associated with the 
multipolar order. In the future, we expect the unprecedent
simplicity of the model and the absence of Monte
Carlo sign problem will allow a direct comparison
between theories, numerics, and experiments on these peculiar doublets.

\emph{Acknowledgments.}---This work is supported by
the Start-up Fund of Fudan University and the National
Thousand-Young-Talents Program of People's Republic of China.
Research at Perimeter Institute is supported by the Government of 
Canada through the Department of Innovation, Science and Economic
Development Canada and by the Province of Ontario
through the Ministry of Research, Innovation and Science.

\bibliography{Refs_Aug}

\appendix

\section{I. Dipole-octupole doublet}

We consider the general wavefunctions of a DO doublet that are
linear superpositions of the $J^z$ states with odd integer multiples
of $3/2$,
\begin{eqnarray}
| \Psi_+\rangle  &=& \sum_{n_1 >0} a_{n_1} |J^z=\frac{3n_1}{2}\rangle
+ \sum_{n_2 < 0} a_{n_2} |J^z = \frac{3n_2}{2}\rangle , 
\\
| \Psi_-\rangle  &=& 
\sum_{n_1 >0} (-)^{\frac{n_1 + 1}{2}} a^{\ast}_{n_1} |J^z = -\frac{3n_1}{2} \rangle
\nonumber  \\
&+& \sum_{n_2 < 0} (-)^{\frac{n_2 +1}{2}} a^{\ast}_{n_2} |J^z = -\frac{3n_2}{2} \rangle ,
\end{eqnarray}
in which $| \Psi_-\rangle $ is simply obtained from $| \Psi_+\rangle $ by a time
reversal operation. Here, both $n_1$ and $n_2$ are
odd integers by definition, and we assume the wavefunctions have been properly 
normalized. Using the definition of the effective spin operator in the main text,
we can relate the effective spin $\tau^{\mu}$ with the total moment $J^{\mu}$ 
as follows
\begin{eqnarray}
\tau^z & \propto & P J^z P,
\label{rel1}
\\
\tau^+ & \propto & P  (J^+)^{3n_1} P \quad \text{  or  } \quad \propto
P  (J^-)^{3|n_2|} P,
\label{rel2}
\\
\tau^- & \propto & P  (J^-)^{3n_1} P \quad \text{  or  } \quad \propto
P  (J^+)^{3|n_2|} P,
\label{rel3}
\end{eqnarray}
where $P=| \Psi_+\rangle \langle \Psi_+| + | \Psi_-\rangle \langle \Psi_-| $
is the projection operator that projects onto the DO doublet manifold.
In Eq.~(\ref{rel2}) and Eq.~(\ref{rel3}), the lowest order in $J^{\pm}$
is $(J^{\pm})^3$. Although the magnetic field couples linearly to $J^{\mu}$,
only $\tau^z$ component survives after we restrict the magnetic field  
coupling to the DO doublet. The octupole moment $\tau^x$, however,
can couple to the magnetic field in the cubic order.

\section{II. Space group symmetry}

\begin{figure}[h]
\centering
\includegraphics[width=4cm]{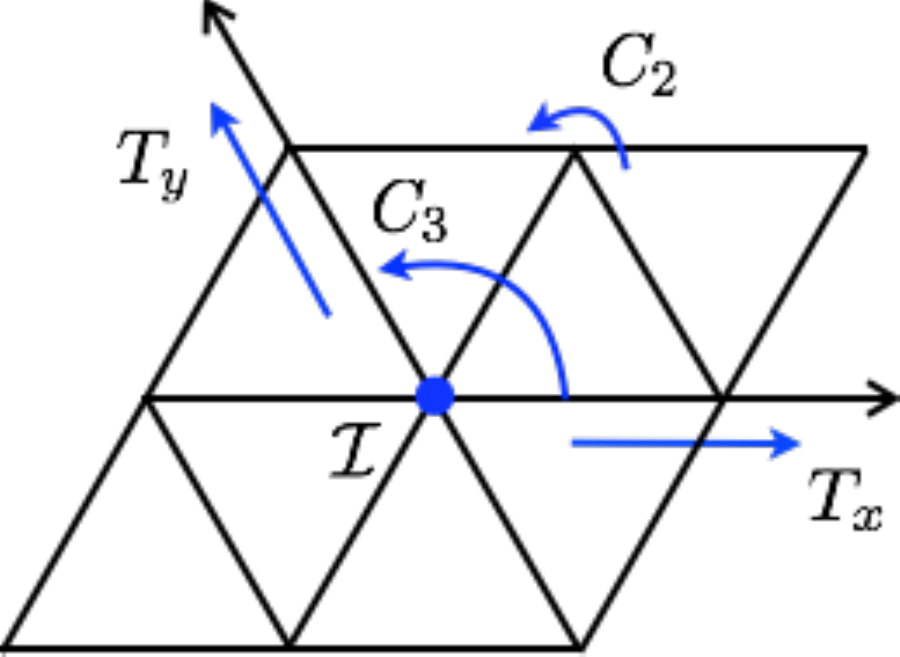}
 \caption{The generators of the space group symmetry for a single
 triangular layer.
}
\label{sfig1}
\end{figure}

As we have explained in the main text, we only need to keep the
space group symmetry generators of the R$\bar{3}$m or P6$_3$mmc
space group. Within the triangular layer,
both R$\bar{3}$m and P6$_3$mmc space groups give the same
list of symmetry generators. As we show in Fig.~\ref{sfig1},
we have the three-fold rotation, $C_3$,
the two-fold rotation, $C_2$, the inversion, $I$,
and two lattice translations, $T_x$ and $T_y$.
Under the symmetry operation, the total moment $J^{\mu}$
transforms as
\begin{widetext}
\begin{eqnarray}
\left\{
\begin{array}{llll}
C_3: & J^z_{\bf r} \rightarrow J^z_{C_3 ({\bf r})}, & J^+_{\bf r} \rightarrow e^{-i\frac{2\pi}{3}}J^+_{C_3 ({\bf r})}, &J^-_{\bf r} \rightarrow e^{i\frac{2\pi}{3}}J^-_{C_3 ({\bf r})},
\\
C_2: & J^z_{\bf r} \rightarrow - J^z_{C_2 ({\bf r})}, & J^+_{\bf r} \rightarrow e^{i\frac{2\pi}{3}}J^-_{C_2 ({\bf r})}, &J^-_{\bf r} \rightarrow e^{-i\frac{2\pi}{3}}J^+_{C_2 ({\bf r})},
\\
I: & J^z_{\bf r} \rightarrow J^z_{I ({\bf r})}, & J^+_{\bf r} \rightarrow J^+_{I({\bf r})}, &J^-_{\bf r} \rightarrow J^-_{I({\bf r})},
\\
T_x: & J^z_{\bf r} \rightarrow J^z_{T_x({\bf r})}, & J^+_{\bf r} \rightarrow J^+_{T_x({\bf r})}, &J^-_{\bf r} \rightarrow J^-_{T_x({\bf r})},
\\
T_y: & J^z_{\bf r} \rightarrow J^z_{T_y({\bf r})}, & J^+_{\bf r} \rightarrow J^+_{T_y({\bf r})}, &J^-_{\bf r} \rightarrow J^-_{T_y({\bf r})},
\end{array}
\right.
\end{eqnarray}
\end{widetext}
Using the relations in Eqs.~(\ref{rel1}-\ref{rel3}),
we obtain the symmetry properties of the pseudospin $\tau^{\mu}$.

\section{III. The transformation for the pseudospin}

In the transformation that we did to eliminate the crossing coupling between
$\tau^y$ and $\tau^z$, we choose the $\theta$ variable such that
\begin{eqnarray}
\sin 2\theta &=& \frac{2J_{yz}}{ [ (J_y - J_z)^2 + (2 J_{yz})^2 ]^{\frac{1}{2}} } \\
\cos 2\theta &=& \frac{J_y-J_z}{ [ (J_y - J_z)^2 + (2 J_{yz})^2 ]^{\frac{1}{2}} },
\end{eqnarray}
and, the new couplings in the reduced model are given as
\begin{eqnarray}
{\mathcal J}_x & = & J_x , \\
{\mathcal J}_y & = & \frac{1}{2} \big[ J_y + J_z + (J_y  - J_z) \cos(2 \theta) 
              \nonumber \\ 
               && \quad\quad + 2 J_{yz} \sin (2 \theta)\big], \\
{\mathcal J}_z & = & \frac{1}{2} \big[ J_y + J_z - (J_y  - J_z) \cos(2 \theta) 
               \nonumber \\ 
               && \quad\quad - 2 J_{yz} \sin (2 \theta) \big].
\end{eqnarray}

\section{IV. Mean field theory in the ferro-octupolar ordered regime}

Starting with the model in Eq.~(\ref{newmodel}), 
we apply mean field decoupling of terms quadratic 
in $T^\mu$ by neglecting their fluctuations,
\begin{equation}
	T^\mu_{\bf r} T^\mu_{{\bf r}'} \to \avg{T^\mu_{\bf r}}  T^\mu_{{\bf r}'} + T^\mu_{\bf r} \avg{T^\mu_{{\bf r}'}} - \avg{T^\mu_{\bf r}} \avg{T^\mu_{{\bf r}'}}.
\end{equation}
For the ferromagnetic order, we can assume a site-independent ansatz, and define $m^\mu \equiv \avg{T^\mu_{\bf r}}$. This gives us the mean-field Hamiltonian,
\begin{eqnarray}
	H_{\rm MF} &=& 6 \sum_{\bf r} \left[ \mathcal{J}_x m^x T^x_{\bf r} + \mathcal{J}_y m^y T^y_{\bf r} + \mathcal{J}_z m^z T^z_{\bf r}\right]
\nonumber \\	
&&	 - h \sum_{\bf r} [\cos \theta \, T^z_{\bf r} + \sin \theta \, T^y_{\bf r}].
\end{eqnarray}
This Hamiltonian can be diagonalized, and $m^\mu$ can be solved self-consistently. For dominant $\mathcal{J}_x$, we may further assume $m^y = m^z = 0$.

At $T=0$ and $h=0$, it is obvious that $m^x = 1/2$. For finite $T$, the self-consist equation is given by $m^x = \frac{1}{2} \tanh{\frac{3m^x}{T}}$. Since $T^x$ does not couple to $h$ linearly, an infinitesimal $h$ would not alter the form of this self-consistent equation. It can be shown that $m^y \sim \frac{\tanh{(3m^x/T)}h}{2 m^x}$ and $m^z \sim \frac{\tanh{(3m^x/T)}h}{2 m^x}$, hence a constant $\chi^{zz}$ below $T_c$.

\section{V. Mean field theory in the antiferro-octupolar ordered regime}

\begin{figure}[h]
\centering
\includegraphics[height=.2\textwidth]{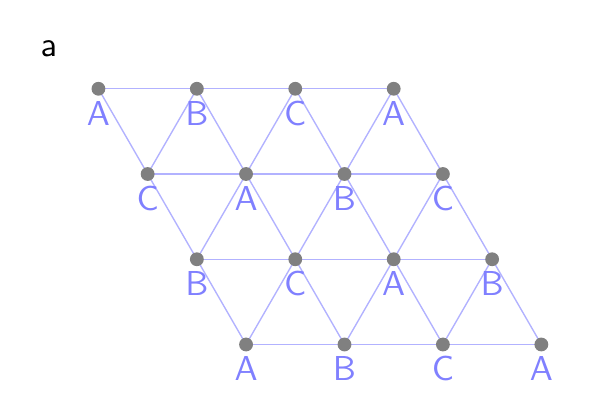}
\includegraphics[height=.2\textwidth]{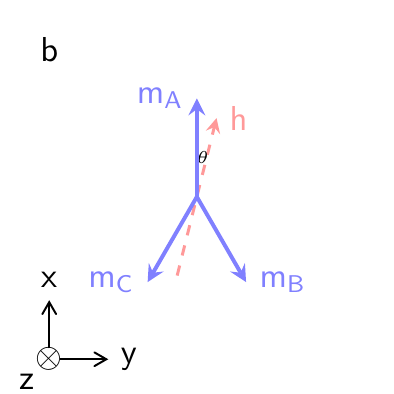}
\includegraphics[height=.2\textwidth]{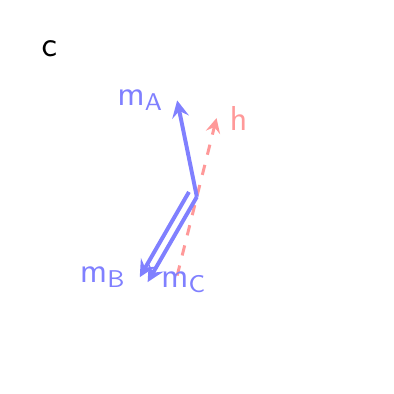}
 \caption{(a) The 3-sublattice structure on the triangular lattice. 
 (b), (c) Two mean-field ansatze of the magnetization vectors for 
 dominant antiferromagnetic $\mathcal{J}_x$. These patterns are 
 obtained by mimicking the supersolid order obtained for the XXZ model~\cite{PhysRevLett.95.127207,PhysRevLett.95.127205,PhysRevLett.112.127203,PhysRevB.91.081104}. 
 The inset is the coordinate system in the pseudospin space. }
\label{sfig2}
\end{figure}

When the dominant exchange is antiferromagnetic, we assume 
a 3-sublattice structure (see Fig.~\ref{sfig2}) for the 
mean-field ansatz, by defining the magnetization 
for each sublattice, ${\bf m}_i = \avg{{\bf T}_i}$, 
where $i ={\text{A, B, C}}$. This assumption is consistent 
with results in the XXZ model. As a result, the mean field 
Hamiltonian reads
\begin{eqnarray}
	H_{\rm MF} &=& 3 \sum_{{\bf r} \in A} \sum_\mu \left[ \mathcal{J}_\mu (m_B^\mu + m_C^\mu) T_{\bf r}^\mu \right] 
\nonumber \\	
&+& 3 \sum_{{\bf r} \in B} \sum_\mu \left[ \mathcal{J}_\mu (m_C^\mu + m_A^\mu) T_{\bf r}^\mu \right]
\nonumber \\
&+& 3 \sum_{{\bf r} \in C} \sum_\mu \left[ \mathcal{J}_\mu (m_A^\mu + m_B^\mu) T_{\bf r}^\mu \right] 
\nonumber \\
&-& h \sum_{\bf r} [\cos \theta \, T^z_{\bf r} + \sin \theta \, T^y_{\bf r}].
\end{eqnarray}
To reduce the number of free parameters, we further constraint 
the magnetization vectors to form patterns depicted in Fig.~\ref{sfig2}b,c. 
The Hamiltonian on each sublattice can now be diagonalized separately, 
and we solve for ${\bf m}_i$ self-consistently. We determine the phase 
diagram by comparing the mean-field ground state energy between the 
two possible patterns of orderings and measuring the suppression 
of $\avg{T^x}$.

\section{VI. Linear spin wave theory}

Our mean field theory gives the magnetization vectors for 
different parameter regimes. Within such phases, there is a 
stable magnetic ordering, therefore spin wave excitations 
are well-defined. Using neutron scattering one can measure 
the spin wave spectrum, as an indirect probe of the ground state.

Suppose the magnetization on site $i$ is given by ${\bf m}_i$, 
we introduce the Holstein-Primakoff representation for the 
pseudospin-$\frac{1}{2}$ operators,
\begin{eqnarray}
	{\bf T}_i \cdot \hat{m}_i &=& \frac{1}{2} - a^\dagger_i a^{\phantom\dagger}_i, \label{eq3}\\
	{\bf T}_i \cdot \hat{z}_i &=& \frac{1}{2}
	  (a^{\phantom\dagger}_i + a^\dagger_i),\label{eq4} \\
	{\bf T}_i \cdot [\hat{m}_i \times \hat{z}_i] &=& \frac{1}{2i}
	  (a^{\phantom\dagger}_i - a^\dagger_i),\label{eq5}
\end{eqnarray}
where $\hat{m}_i$ is the unit vector parallel to ${\bf m}_i$, 
and $\hat{z}_i$ is a unit vector perpendicular to $\hat{m}_i$. 
In this representation, the Bloch Hamiltonian has the form
\begin{equation}
H_{\text{HP}} = \sum_{{\bf k}\in \text{BZ}'} (A^{\dagger}_{\bf k}, A_{-{\bf k}}^{\phantom\dagger})
				\left(
					\begin{array}{ll}
						F_{\bf k} & G^{\dagger}_{\bf k} \\
						G_{\bf k} & F_{-{\bf k}}
					\end{array}
				\right)
				\left(
					\begin{array}{l}
						A_{\bf k}\\
						A^{\dagger}_{-{\bf k}}
					\end{array}
				\right),
\end{equation}
where $A_{\bf k}
= (a_{1{\bf k}}, \ldots, a_{n{\bf k}})$ is the vector
of boson annihilation operators, the
subindices $1 \ldots n$ label
the $n$ sublattices of the magnetic unit cell,
and BZ$'$ is the magnetic Brioullin zone.
$F_{\bf k}$ and $G_{\bf k}$ are $2 \times 2$
matrices and depend on the mean field magnetizations.
The Bloch Hamiltonian is diagonalized by the standard Bogoliubov 
transformation, giving the spectrum of Holstein-Primakoff 
bosons plotted in Fig.~\ref{fig1} and Fig.~\ref{fig3}.

\end{document}